\documentclass{elsart3}
\usepackage[dvips]{graphicx}

\def\degree{{${}^\circ$}}

\begin{document} 

\renewcommand{\thefootnote}{\fnsymbol{footnote}}

\onecolumn

\begin{flushright}
{\small
SLAC--PUB--11833\\
May, 2006\\}
\end{flushright}

\vspace{.8cm}

\begin{center}
{\bf\large   
GLAST Tracker
\footnote{This work was supported in part by the U.S. Department of Energy under Grant DE-AC02-76SF00515 and 22428-443410, and in part by the Agenzia Spaziale Italiana (ASI).}}

\vspace{1cm}

Hiroyasu Tajima

On behalf of GLAST Tracker Team
\medskip

{Stanford Linear Accelerator Center, Stanford, CA 94309-4349, USA}

\end{center}

\vfill

\begin{center}
{\bf\large   
Abstract }
\end{center}

\begin{quote}
The Large Area Telescope (LAT) on board the Gamma-ray Large-Area Space Telescope (GLAST) is a pair-conversion gamma-ray detector designed to explore the gamma-ray universe in the 20~MeV--300~GeV energy band.
The Tracker subsystem of the LAT will perform tracking of electron and positrons to determine the origin of the gamma-ray.
The design and performance of the GLAST LAT Tracker are described in this paper.
\end{quote}

\bigskip


\vfill

\begin{center} 
{\it Invited talk at} 
{\it Vertex 2005, Chuzenji Lake, Nikko, Japan, November 7--November 11, 2005} \\
{\it To be published in} 
{\it Nuclear Instruments and Methods A}\\



\end{center}

\twocolumn
\clearpage

\section{Introduction}

The Large Area Telescope (LAT) of the Gamma-ray Large-Area Space Telescope (GLAST) mission \cite{Atwood94,Gehrels99} is a pair-conversion gamma-ray detector similar in concept to the previous NASA high-energy gamma-ray mission EGRET on the Compton Gamma-Ray Observatory \cite{Thompson93}.  
High energy (20~MeV--300~GeV) gamma-rays convert into electron-positron pairs in one of 16 layers of tungsten foils.
The charged particles pass through up to 36 layers of position-sensitive detectors interleaved with the tungsten, the ``tracker," leaving behind tracks pointing back toward the origin of the gamma ray.  
After passing through the last tracking layer they enter a calorimeter composed of bars of cesium-iodide crystals read out by PIN diodes.
The calorimeter furnishes the energy measurement of the incident gamma ray.
A third detector system, the anti-coincidence detector (ACD), surrounds the top and sides of the tracking instrument.
It consists of panels of plastic scintillator read out by wave-shifting fibers and photo-multiplier tubes and is used to veto charged cosmic-ray events such as electrons, protons or heavier nuclei.  

In the LAT the tracker and calorimeter are segmented into 16 ``towers," which are covered by the ACD and a thermal blanket and meteor shield.
An aluminum grid supports the detector modules and the data acquisition system and computers, which are located below the calorimeter modules.
The LAT is designed to improve upon EGRET's sensitivity to astrophysical gamma-ray sources by well over a factor of 10.
That is accomplished partly by sheer size, but also by use of state-of-the-art particle detection technology, such as the silicon-strip detectors\cite{Ohsugi99} used in the tracker system.

\section{GLAST LAT Tracker}

\begin{figure}[tbh]
   \begin{center}
   \includegraphics[width=4.0cm]{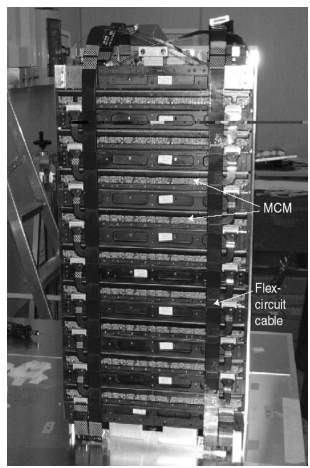}
   \end{center}
   \caption
   { \label{fig:module} Inverted view of one tracker module, with a sidewall removed. Nine MCMs and 2 flex-circuit cables are visible. }
   \end{figure}

Each of the 16 tracker modules is composed of a stack of 19 ``trays," as can be seen Fig.~\ref{fig:module}.
A tray is a stiff, lightweight carbon-composite panel with silicon-strip detectors (SSDs) bonded on both sides, with the strips on top parallel to those on the bottom.
Also bonded to the bottom surface of all but the 3 lowest trays, between the panel and the detectors, is an array of tungsten foils, one to match the active area of each detector wafer.
The thickness of the tungsten foil is 3\% radiation length for the upper 12 trays (light-converter trays), 18\% radiation length for the next 4 trays (thick-converter trays). 
The last 3 trays do not have tungsten foils.
Each tray is rotated 90\degree\ with respect to the one above or below.
The detectors on the bottom of a tray combine with those on the top of the tray below to form a 90\degree\ stereo x,y pair with a 2~mm gap between them, and with the tungsten converter foils located just above.

Each front-end electronics multi-chip module (MCM) supports the readout of 1536 silicon strips.
It consists of a single printed wiring board (PWB) upon which are mounted 24 64-channel amplifier-discriminator ASICs (GTFE), two digital readout-controller ASICs (GTRC), the right-angle interconnect, bias and termination resistors, decoupling capacitors, resettable fuses, and two nano-connectors.
Each nano-connector plugs into a long flex-circuit cable, each of which interfaces 9 MCMs to the data-acquisition electronics located below the calorimeter in the Tower Electronics Module (TEM).
Thus on each of the 4 sides of a tracker module one finds 9 readout boards to support 9 layers of silicon-strip detectors, which send their data to the TEM via two flex-circuit cables (see Fig.~\ref{fig:module}).

Each channel in the GTFE has a preamplifier, shaping amplifier, and discriminator similar, although not identical, to the prototype circuits described in \cite{Johnson98}.  
The amplified detector signals are discriminated by a single threshold per GTFE chip; no other measurement of the signal size is made within the GTFE.
The GTFE chips are arranged on the MCM in 4 groups of 6.
Each group reads out one SSD ``ladder," which consists of 4 SSDs connected in series to yield strips of about 36 cm effective length.

All communication with the TEM passes through the GTRC chips, which in turn relay commands and data to and from the GTFE chips.
Event data and trigger primitives flow from the GTFE chips into one or the other of the GTRC chips by passing through one GTFE chip after another.
This scheme was chosen over the use of a common bus in order to avoid the possibility of a single malfunctioning chip pulling down the entire bus.
Concern that in the chosen scheme a single bad chip could block the flow of data is mitigated by the left-right redundancy described below.

Each GTFE chip has two command decoders, one that listens to the left-hand GTRC, and a second that listens to the right-hand GTRC.
Each GTFE also has two output data shift registers, one that moves data to the left, and a second that moves data to the right.
Trigger information is formed within each GTFE chip from a logical OR of the 64 channels, of which any arbitrary set can be masked.
The OR signal is passed to the left or right, depending on the setting of the chip, and combined with the OR of the neighbor, and so on down the line, until the GTRC receives a logical OR of all non-masked channels in those chips that it controls.
This ``layer-OR" trigger primitive initiates in the GTRC a one-shot pulse of adjustable length, which is sent down as a ``trigger request" to the TEM for trigger processing.
In addition, a counter in the GTRC measures the length of the layer-OR signal (time-over-threshold) and buffers the result for inclusion in the event data stream.
Upon receipt of a ``trigger acknowledge", each GTFE chip latches the status of all 64 channels into one of 4 internal event buffers, as specified by the 2-bit trigger code.
A 64-bit mask, which is separate from the trigger mask mentioned above, can be used to mask any subset of channels from contributing data, as may be necessary in case of noisy channels.

\section{Performance}

We have completed fabrication of all 16 flight modules of GLAST LAT Tracker.
In compliance with NASA's rigorous requirement verification program, every tracker flight module was tested to confirm its performance in detail.
In this section, we describe the test results of major performance quantities.

Average hit efficiency must be better than 98\% to assure adequate tracking efficiency.
The number of bad channels is monitored throughout all construction stages to assure good hit efficiency.
We have three types of bad channels: dead, hot and disconnected channels.
A dead channel is one in which we receive no response from test charge injections.
On average, 0.05\% of channels are identified as dead, and no layer has more than 2\% of hot channels.
A hot channel is defined in terms of data volume and trigger occupancy.
The average noise occupancy is required to be less than $5\times10^{-5}$ to suppress useless data from using up the downlink between satellite and ground.
In each tracker module, the highest occupancy channels are identified as hot and masked from data collection until the average occupancy is below the above requirement. Trigger occupancy is required to be less than 8\% for each layer, which empirically corresponds to an average strip occupancy of $1.5\times10^{-5}$ for a nominal trigger pulse width of 1.6~$\mu$s.
The highest occupancy channels are identified as hot and masked from the trigger until the average occupancy in each layer becomes below the above requirement.
On average, 0.03\% of channels are identified as hot, and no layer has more than 2\% hot channels.

\begin{figure}[tbh]
   \begin{center}
   \includegraphics[height=4.1cm]{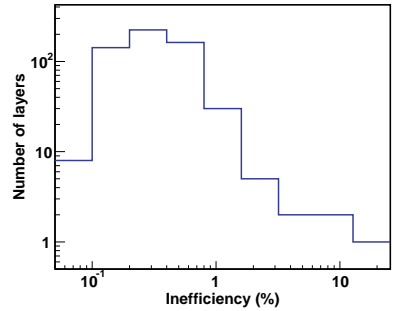}
   \end{center}
   \caption{ \label{fig:eff} Distribution of hit efficiency for each layer estimated using cosmic-ray muon tracks. }
   \end{figure}

Disconnected channels are by far the most common type of bad channel and are
due to cracked traces on the pitch adapter or broken wire-bonds between ASIC
and SSD or between SSDs. Cracked pitch adapter traces were caused by bending
of the brittle nickel plating required for wire bonding but were kept at an
acceptable level by controlling the plating thickness and type
(electrolytic).  Assembly tolerances did not allow the plating to be
eliminated from the bend region.  Broken wire-bonds were caused by
delamination of the encapsulation materials during thermal cycle tests in
the early phase of the construction and account for most of the disconnected
channels. Two separate problems contributed to the delamination issues.
Delamination of the epoxy encapsulation on the MCMs was due to silicone
contamination of the Kapton surface where the encapsulation was to adhere.
That problem was identified and corrected about 1/3 of the way into the
production.   Delamination of the Nusil encapsulation on the SSDs resulted
from the thermal-expansion mismatch between the tungsten foils and the
carbon-composite trays.  The thermal distortion was large enough for the
thick-converter trays (but not the light-converter trays) to cause
delamination of some of the Nusil encapsulation from the SSD surfaces.  That
problem was solved by eliminating the encapsulation from ladders on
heavy-converter trays, but only after the first two towers were fabricated.
The effects on the second tower were minimized by reducing the thermal-cycle
temperature range (which required modification of the thermal safety system
of the instrument).
On average, 0.3\% of the channels are identified as disconnected and 12 out of 576 layers have more than 2\% disconnected channels.
Note that having more than 2\% bad channels does not directly corresponds to having more than 2\% hit inefficiency, since tracks often leave two or more hits per layer, and one of them can be detected even when the others are dead.
After module construction was completed, the hit efficiency was measured using cosmic-ray muon tracks.
Fig.~\ref{fig:eff} shows the distribution of hit efficiency for each layer.
Average hit efficiency is found to be 99.6\%, which is above the requirement of 98\% by a wide margin, and 99\% of the layers have more than 98\% hit efficiency.

\begin{figure}[tbh]
   \begin{center}
   \includegraphics[height=4.1cm]{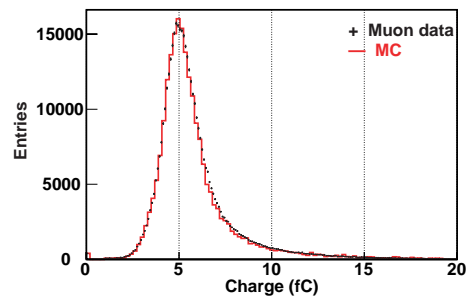}
   \end{center}
   \caption{ \label{fig:tot} Comparison of the TOT distribution of cosmic-ray muon tracks for data (points) and a Monte Carlo simulation sample (histogram). }
   \end{figure}

The TOT (time-over-threshold) provides a maximum pulse height measurement for each GTRC (a half layer).
The TOT gain varies from channel to channel by 30\% rms due to dispersion of the shaper fall time. (The peaking time is relatively uniform.)
The relative gain of the TOT is calibrated using test pulses.
Absolute TOT calibration for each GTFE is performed using cosmic-ray muon tracks, which also set the absolute scale for the GTFE charge injector.
This absolute charge injection scale is used for subsequent calibration of discriminator thresholds and measurements of trigger time walk.
Fig.~\ref{fig:tot} compares the TOT distribution of cosmic-ray muon tracks for data after calibration (points) and a Monte Carlo simulation sample (histogram).
The distributions agree with each other very well.
The average correction factor for the charge injection scale is measured to be 1.13 with an rms variation of 8.4\%.

\begin{figure}[tbh]
   \begin{center}
   \includegraphics[height=4.1cm]{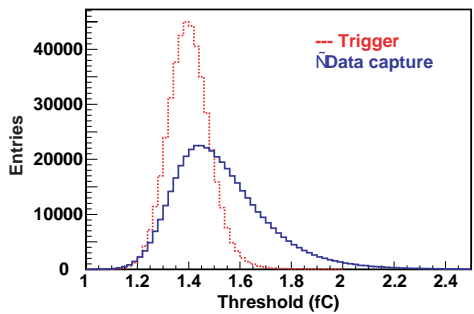}
   \end{center}
   \caption{ \label{fig:threshold} Threshold distributions for the trigger (dotted) and data latch (solid). }
   \end{figure}

The discriminator threshold is nominally set at 1.4~fC (approximately 1/4 of a minimum ionizing particle signal for 400~$\mu$m thick Si.)
Due to channel gain variation, the threshold dispersion within each GTFE is 5.2\%, while the threshold dispersion between GTFEs after calibration is 2.7\%, due to granularity of the threshold DAC. (The nominal threshold value is around 30 DAC counts.  Therefore we expect a 3\% accuracy in the threshold setting.)
This performance meets our specification of 10\%, designed to minimize the effect on the trigger jitter.
As described above, the hit data are latched later than the trigger signals.
At the nominal trigger delay and coincidence window width, we expect the data latch to occur 1~$\mu$s after the shaper peak. ($\sim$2~$\mu$s after the tracker trigger request.)
This means that the ``effective threshold" for the data latch is somewhat higher than the threshold for the trigger.
The threshold dispersion for the data latch is also larger, due to variation of the shaper fall time.
Total threshold dispersion for the data latch is measured to 12.0\%, of which 8\% comes from channel-to-channel variations and 7\% comes from GTFE-to-GTFE variations.
Fig.~\ref{fig:threshold} shows threshold distributions for the trigger (dotted) and data latch (solid).

\begin{figure}[tbh]
   \begin{center}
   \includegraphics[height=4.1cm]{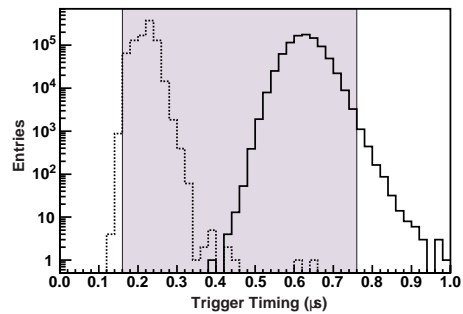}
   \end{center}
   \caption{ \label{fig:time-walk} Trigger timing distributions for the injection charges of 2.5~fC (solid) and 20--30~fC (dotted). }
   \end{figure}

The trigger time walk needs to be minimized to suppress chance coincidences with the ACD veto.
After the threshold DACs were calibrated, the trigger time walk between injection charges of 2.5~fC and 20--30~fC was measured.
Figure~\ref{fig:time-walk} shows the trigger timing distributions for injection charges of 2.5~fC (solid) and 20--30~fC (dotted).
We find that 99.8\% of the channels are within the 0.6~$\mu$s requirement, which is indicated by the hatched region in Fig.~\ref{fig:time-walk}

\section{Conclusions}
We have completed fabrication of all 16 flight modules of the GLAST LAT Tracker.
Detailed requirement verification tests have demonstrated that all of the tracker modules are in compliance with the specification required to carry out the science objectives of the GLAST mission.



\end{document}